\documentclass[trackchanges,twocolumn]{aastex701}
\usepackage{verbatim}

\begin{document}

\title{
Searching for Gamma Ray Bursts associated with the Second CHIME/FRB Catalog}

\author[orcid=0009-0005-0170-192X]{Yifang Liang}
\affiliation{Purple Mountain Observatory, Chinese Academy of Sciences, Nanjing 210008, China}
\affiliation{School of Astronomy and Space Sciences, University of Science and Technology of China, Hefei 230026, China}
\email[]{yfliang@pmo.ac.cn}

\author[orcid=0000-0001-5931-2381]{Ye Li$^\dag$}
\affiliation{Purple Mountain Observatory, Chinese Academy of Sciences, Nanjing 210008, China}
\affiliation{State Key Laboratory of Radio Astronomy and Technology, Purple Mountain Observatory, Chinese Academy of Sciences, 10 Yuanhua Road, Nanjing 210023, China}
\email[show]{yeli@pmo.ac.cn}

\author[orcid=0000-0003-3635-5375]{Bao Wang}
\affiliation{Purple Mountain Observatory, Chinese Academy of Sciences, Nanjing 210008, China}
\affiliation{School of Astronomy and Space Sciences, University of Science and Technology of China, Hefei 230026, China}
\email[]{bwang@pmo.ac.cn}

\author[]{Xuan Yang}
\affiliation{Purple Mountain Observatory, Chinese Academy of Sciences, Nanjing 210008, China}
\email[]{yangxuan@pmo.ac.cn}

\author{Yuan-Pei Yang}
\affiliation{South-Western Institute for Astronomy Research, Yunnan University, Kunming 650504, China}
\email[]{ypyang@ynu.edu.cn}

\author[orcid=0000-0002-6299-1263]{Xuefeng Wu}
\affiliation{Purple Mountain Observatory, Chinese Academy of Sciences, Nanjing 210008, China}
\affiliation{State Key Laboratory of Radio Astronomy and Technology, Purple Mountain Observatory, Chinese Academy of Sciences, 10 Yuanhua Road, Nanjing 210023, China}
\email[]{xfwu@pmo.ac.cn}

\correspondingauthor{Ye Li}

\begin{abstract}
Fast radio bursts (FRBs) and gamma-ray bursts (GRBs) are both linked to compact-object activity, yet their possible connection remains unclear. Here we perform a systematic search for spatial and temporal associations between FRBs in the second CHIME/FRB catalog and Swift GRBs.
Instead of using the positional ellipses reported in the catalog, the full CHIME localization probability maps are adopted for spatial cross-matching. This yields 130 candidate pairs and increases the number of spatially consistent matches by a factor of several.
Applying a distance-consistency criterion based on DM-inferred FRB redshifts and GRB distances inferred via the Amati relation reduces the sample to 37 pairs, including 26 GRB-preceding-FRB candidates (24 LGRB--FRB and 2 SGRB--FRB).
Monte Carlo simulations show that the overall excess of associations is not statistically significant, and the distribution of matches across localization confidence levels is consistent with random expectations. These pairs are therefore not claimed as secure associations, but are used to constrain a possible subdominant FRB--GRB connection.
These results place constraints on any FRB--GRB connection and highlight the need for improved localization and larger samples.

\end{abstract}


\section{Introduction} \label{sec:intro}

Fast Radio Bursts (FRBs) are luminous ($\sim10^{44}$~erg~s$^{-1}$), millisecond-duration radio transients of extragalactic origin \citep{2007Lorimer,2013Thornton}. Despite the rapidly growing sample---with more than 5000 events reported to date---their physical origin remains an open question. The all-sky event rate of FRBs, estimated to be $\sim10^{3}$~day$^{-1}$ \citep{Amiri_2021}, is comparable to or higher than that of several classes of stellar end-of-life explosions. This suggests that most FRB sources are capable of repeating activity, even though only a small fraction have been observed to do so \citep{Ravi2019NatAs,2023chime,2025ApJBeniamini}.
Recent observational progress, led by the Canadian Hydrogen Intensity Mapping Experiment (CHIME) \citep{2018chimeintro}, has drastically expanded the FRB population. In particular, the second CHIME/FRB Catalog \citep{chime2ndcatalog} 
provides a sample of more than 4000 FRBs, making it well suited for statistical studies.
Although CHIME features relatively coarse single-event localizations compared to interferometric arrays, it delivers comprehensive pixelized probability maps that fully describe positional uncertainties, making these data ideally suited for population-scale association frameworks.

A growing body of evidence suggests that at least a fraction of FRBs originate from magnetars \citep{ZhangB_2020,ZhangB_2024,xd2021,Platts2019}. The repeating source FRB~121102, located in a low-metallicity star-forming dwarf galaxy, points to a young and highly active progenitor \citep{Scholz2016,2017Beloborodov,Marcote2017,Tendulkar2017}. More directly, an FRB-like burst was detected from the Galactic magnetar SGR~1935+2154, accompanied by a coincident X-ray flare detected by high-energy observatories \citep{CHIME2020nature,2020Natur.587...59B,LickNatAs,RidnaiaNA}, demonstrating that magnetars can produce coherent radio bursts with properties analogous to extragalactic FRBs. Population studies further show that FRBs preferentially occur in star-forming galaxies, consistent with progenitors linked to recent stellar evolution \citep{Liye2020ApJ,Bochenek_2021,2022Bhandari,2023Gordon,2024ApJBhardwaj}. These results support the idea that young magnetars formed in energetic transients may be a main channel for FRB production.

Gamma-ray bursts are among the most energetic transients in the universe and are broadly classified into short GRBs (SGRBs) and long GRBs (LGRBs) based on their duration $T_{\rm 90}$ \citep{Zhang_2018}. SGRBs are generally associated with compact binary mergers, while LGRBs are linked to the collapse of massive stars. Both channels are capable of producing highly magnetized neutron stars under certain conditions. 
Several theoretical models predict FRB--GRB connections on distinct timescales. For quasi-simultaneous emission, pre-merger magnetospheric activity in SGRB progenitors could produce FRB emission shortly before or during the merger \citep{Totani2013,2014A&AFalcke,2017ApJMetzger,Yamasaki2018,WangFY_2020}. Additionally, interaction between the GRB jet and its surrounding medium may generate contemporaneous radio emission in either SGRBs or LGRBs \citep[see][]{Metzger2019}. For delayed emission, long-lived magnetar remnants formed in either LGRBs or SGRBs could power FRB activity with time delays more than months to years, once the surrounding ejecta becomes transparent \citep{2014ApJZhang,Murase2016,Rowlinson2013,Rowlinson2014,Metzger2019,Margalit2019,ZhangB_2020}. These diverse predictions motivate a systematic search across a broad temporal range.

To test these scenarios, numerous campaigns have searched for spatial and temporal coincidences.
While a few candidate associations have been discussed in the literature \citep[e.g.,][]{Wangxg_2020,lumx2023,LiY2025}, no definitive physical association has been confirmed to date. These efforts have nevertheless placed important statistical constraints on prompt-time connections \citep[e.g.,][]{Guidorzi2019,Guidorzi2020,Curtin2023} as well as long-term non-simultaneous emissions \citep[e.g.,][]{Men2019,Madison2019,Patricelli2024,Dongy2025,chenhh2026}.
However, differing localization definitions across datasets often challenge the consistent evaluation of these statistical limits.
Recently, \citet{Curtin2024} carried out a systematic cross-matching between the second CHIME/FRB Catalog sample and GRBs detected by \textit{Swift} and \textit{Fermi} satellites,
focusing on near-simultaneous pairs with a $\pm$ 1-week temporal window and a 3-sigma localization criterion. Their analysis provides valuable constraints on possible prompt-time SGRB-FRB connections. Building on this work, we extend the search to broader temporal ranges (including delayed emission scenarios) and adopt a fully probabilistic spatial matching method using the CHIME pixelized localization maps.

In this work, we revisit the FRB--GRB association problem using the same second CHIME/FRB Catalog but implementing a significantly more rigorous, self-consistent statistical framework. 
Instead of using a simple angular separation cut, we adopt a probabilistic spatial association criterion based on the full CHIME pixelized localization maps \citep{LiuFRB&SN2025}. 
Crucially, we first identify spatially consistent pairs without imposing a narrow temporal window, and then apply simple and physically motivated temporal classification. This allows us to expand our search to a much broader non-simultaneous parameter space--including delays of months to years--thereby unlocking new scientific possibilities from the exact same dataset.

The structure of this paper is as follows. In Section~\ref{sec:sample}, we describe the FRB and GRB samples used in this study. Section~\ref{sec:method} presents our spatial matching method and the criteria for temporal and redshift consistency. The results are given in Section~\ref{sec:result}, followed by a discussion in Section~\ref{sec:discuss}. Finally, we summarize our conclusions in Section~\ref{sec:con}.
We adopt a standard $\Lambda$CDM cosmology with $H_0 = 70$ km s$^{-1}$ Mpc$^{-1}$ and $\Omega_m = 0.3$.

\section{Samples}\label{sec:sample}
\subsection{FRB Sample}
We use the second CHIME/FRB catalog \citep{chime2ndcatalog}, which contains a total of 4539 FRBs, including both one-off events and repeating sources. Among them, 981 bursts are associated with 83 repeating sources, while the remaining 3558 events are classified as apparently non-repeating bursts.

The catalog provides sky positions for all bursts, along with associated localization uncertainties. In addition to the commonly used central coordinates and their elliptical errors, CHIME also releases localization probability maps for each FRB. These maps encode the full positional probability distribution on the sky, which can be highly non-Gaussian and irregular due to the instrument beam response. Among them, the localizations of 138 FRBs were replaced with more precise measurements at the $\sim 10^{\prime\prime}$ level, for which channelized raw voltage (``baseband'') data are available \citep{Michilli2021, Michilli2023, chime2024updating}.

For this study, we select 4513 FRBs for which reliable localization maps are available and suitable for cross-matching analysis. For repeating FRBs, we perform the matching at the burst level, because each burst has its own arrival time and the temporal ordering between the GRB and FRB is part of our selection. This treatment is used to preserve the timing information, and does not imply that bursts from the same repeater are independent sources.
The dispersion measure (DM) of each burst is used as a proxy for distance, enabling an approximate inference of the FRB redshift, as described in Section~\ref{sec:method}.

\subsection{GRB Sample}
The GRB sample is constructed from the \textit{Swift} GRB catalog\footnote{\url{https://swift.gsfc.nasa.gov/archive/grb_table.html/}}, covering the period from 2005 to 2025. 
To minimize additional positional uncertainties and chance coincidences, we restrict the sample to GRBs with arcsecond-precision localization obtained from either the X-Ray Telescope (XRT) or the Ultraviolet/Optical Telescope (UVOT).

Among these GRBs, 464 events have spectroscopically measured redshifts reported in the literature. For the remaining GRBs without direct redshift measurements, redshifts are estimated using the empirical Amati relation, as described in Section~\ref{sec:method}.

\section{Methodology}\label{sec:method}

\begin{figure}
    \centering
    \includegraphics[width=0.95\linewidth]{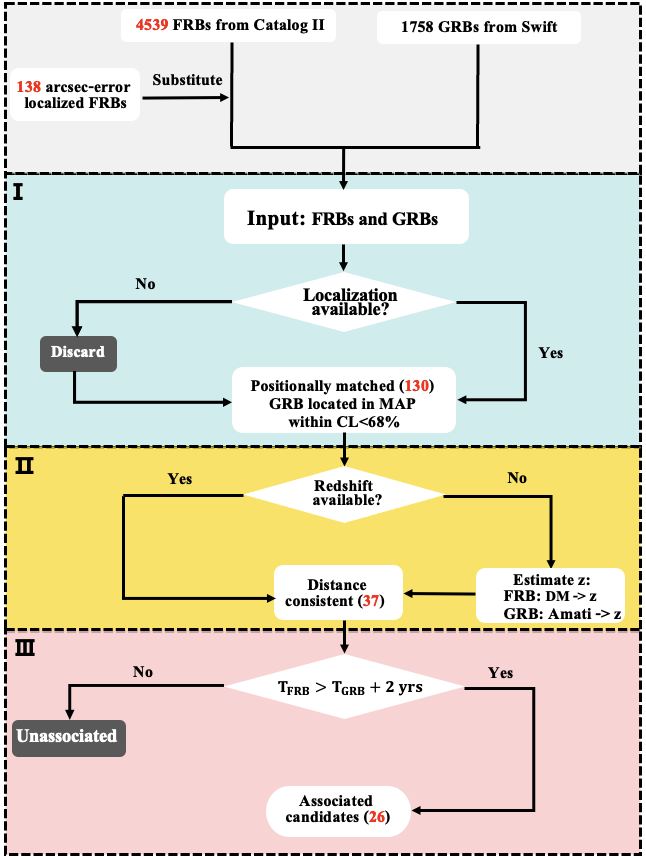}
    \caption{The workflow of the FRB--GRB association procedure. The input consists of FRB and GRB samples, which are processed through spatial and temporal selection steps to identify potential associations. The procedure evaluates positional consistency within localization uncertainties and applies additional constraints (e.g., time delay and redshift consistency) to obtain the final matched sample.
    }
    \label{fig:flowchart}
\end{figure}

We identify candidate FRB--GRB associations through a multi-stage selection procedure that incorporates spatial, redshift, and temporal information. The spatial localization provides the primary constraint, while the redshift and temporal criteria are applied as additional consistency checks. The overall workflow is illustrated in Figure~\ref{fig:flowchart}.

\subsection{Localization-based Matching}\label{sec:method.1}

Instead of adopting a simple angular separation criterion, we utilize the full localization probability information provided by the CHIME/FRB localization maps. 
For each FRB, the CHIME pipeline produces a sky map characterized by a $\Delta\chi^2$ surface. This surface is empirically calibrated using a sample of localized pulsars, allowing $\Delta\chi^2$ values to be mapped to confidence level (CL) \citep{chime2ndcatalog}. Specifically, the CL at a given sky position can be interpreted as the fraction of the total localization probability contained within regions of higher likelihood.

For each FRB--GRB pair, we evaluate the CL at the GRB sky position by interpolating the FRB localization map. Only GRBs located within the valid footprint of the map are considered. A pair is defined as spatially consistent if the GRB position lies within the 68\% region (i.e., CL $< 0.68$) of the FRB localization.

We perform a global search over all FRBs and GRBs, identifying all candidate pairs that satisfy this spatial criterion.

\subsection{Redshift Consistency}\label{sec:redshift_consistency}

For spatially consistent FRB--GRB pairs, we further apply a consistency check based on their inferred redshifts. For FRBs, redshifts are inferred from dispersion measures (DM), while for GRBs, we use spectroscopic or photometric redshifts when available\footnote{\url{https://swift.gsfc.nasa.gov/archive/grb_table.html/}}, and otherwise adopt the Amati relation to estimate redshifts.

We emphasize that these redshift estimates are subject to substantial uncertainties.
Therefore, this step is used as a broad consistency criterion to exclude clearly incompatible pairs, rather than a precise distance matching condition.

\subsubsection{FRB Redshift Estimation from Dispersion Measure}
\label{sec:frb_z}

For FRBs without host galaxy redshift measurements, we infer their redshifts from the observed DM based on the Bayesian inference method \citep{Tang2023ChPhC}. The total DM is decomposed as
\begin{equation}
    \mathrm{DM}_{\rm obs} = \mathrm{DM}_{\rm ISM}^{\rm MW}
    + \mathrm{DM}_{\rm halo}^{\rm MW}
    + \mathrm{DM}_{\rm IGM}(z)
    + \frac{\mathrm{DM}_{\rm host}}{1+z}.
\end{equation}
The interstellar medium (ISM) of the Milky Way contribution $\mathrm{DM}_{\rm ISM}^{\rm MW}$ is estimated using the NE2001 model \citep{Ne2001}. The contribution of the Milky Way halo spans a wide range, from $\sim 20$ to $\sim 100\,\mathrm{pc\ cm^{-3}}$ \citep{Prochaska2019,Cook2023,Hoffmann2026,Liuyang2026}. As a fiducial value, we adopt $\mathrm{DM}_{\rm halo}^{\rm MW} \sim 50\,\mathrm{pc\ cm^{-3}}$ \citep{Macquart_2020}.

The intergalactic medium (IGM) contribution $\mathrm{DM}_{\rm IGM}(z)$ is computed under a flat $\Lambda$CDM cosmology, with stochastic fluctuations modeled as $\sigma_{\rm IGM} \propto z^{-1/2}$. The host-galaxy contribution $\mathrm{DM}_{\rm host}$ is described by a log-normal distribution.
For details, please refer to \citep{Tang2023ChPhC}.

For each FRB, we construct a likelihood function $\mathcal{L}_{\rm FRB}(z)$ and sample the posterior distribution using a Markov Chain Monte Carlo (MCMC) method with a flat prior over $0.001 < z < 4$. The median and 16th–84th percentiles are adopted as the inferred redshift and uncertainty. The estimated FRB redshifts have a median fractional uncertainty of $f_z=(z_{84}-z_{16})/(2z_{50})\sim 35\%$, reflecting the stochastic IGM contribution, the uncertain host-galaxy DM, and line-of-sight variations.

\subsubsection{GRB Redshift Estimation from the Amati Relation} \label{sec:grb_redshift}

For GRBs without spectroscopic or photometric redshifts, we estimate their pseudo-redshifts using the Amati relation \citep{2002Amati,Amati2006,Amati2008}, which connects the rest-frame peak energy $E_{\rm p,i}$ and the isotropic-equivalent energy $E_{\rm iso}$.
We use the \citet{Amati2008} calibration,
\begin{equation}
    \log_{10} E_{\rm p,i} = 2.04 + 0.53\log_{10}(\frac{E_{\rm iso}}{10^{52}~{\rm erg}}),
\end{equation}

For a trial redshift $z$, we calculate
\begin{equation}
E_{\rm iso}(z) =
\frac{4\pi d_L^2(z)}{1+z} \, S_{\rm obs} \, k(z),
\end{equation}
where $d_L(z)$ is the luminosity distance, $S_{\rm obs}$ is the observed fluence, and $k(z)$ converts the measured observer-frame bandpass to the rest-frame 1--10,000 keV band. The $k$-correction is calculated from the spectral model used for the fluence measurement.

For Band spectra, we use the reported photon indices and $E_{\rm p}$. For cutoff power-law (CPL) spectra, we use $N(E)\propto E^{\alpha_{\rm cpl}}\exp(-E/E_{\rm cut})$, with $E_{\rm p}=(2+\alpha_{\rm cpl})E_{\rm cut}$. For bursts without measured redshifts, we first use the \textit{Fermi}/GBM spectral catalog when a matched GBM trigger is available. We adopt the Band fit whenever it is reported, and otherwise the CPL fit, since both models provide a measured $E_{\rm p}$ \citep{FermiGBMCatalog}. If no usable GBM spectrum is available, we search the \textit{Swift}/BAT and GCN records. For BAT bursts with only a simple power-law (PL), $E_{\rm p}$ is not directly measured because of the narrow BAT bandpass. In these cases, we estimate $E_{\rm p}$ from the empirical relation of \citet{Sakamoto2009},
\begin{equation}
    \log_{10}\left(E_{\rm p}/{\rm keV}\right)=3.258+0.829\,\alpha_{\rm pl},\quad -2.3<\alpha_{\rm pl}<-1.3,
\end{equation}
 We use these estimates only when the photon index falls within the calibrated range. These PL-based values are empirical estimates of $E_{\rm p}$, not direct measurements.

The likelihood is written as
\begin{equation}
\mathcal{L}(z) \propto \exp\left[-\frac{1}{2}\frac{\left(\log E_{\rm p,i} - \mu(E_{\rm iso})\right)^2}{\sigma_{\rm A}^2}\right],
\end{equation}
where $\mu$ represents the expected value of the best-fit correlation from the Amati relation, $\sigma_A$ includes the intrinsic scatter (0.17 dex, following \citealt{Amati2008}) and the measurement uncertainty in fluence and $E_{\rm p}$.

The posterior is sampled with a prior uniform in $\log z$ over $0.01 < z < 10$. We use the median as the pseudo-redshift, and the 16th--84th percentile range as its uncertainty. We estimate a pseudo-redshift only when the prompt data provide both $S_{\rm obs}$ and $E_{\rm p}$. This includes measured Band/CPL spectra, and PL spectra whose photon index lies within the \citet{Sakamoto2009} range. If $E_{\rm p}$ is unconstrained or no prompt spectral record is available, we do not assign a GRB pseudo-redshift.

For the GRB pseudo-redshifts, the median fractional uncertainty is $f_z\sim 67\%$. Given these large uncertainties, we require overlap between the FRB and GRB 16th--84th percentile redshift intervals. If the GRB has a precisely measured redshift, we require it to lie within the FRB interval.

\subsection{Temporal Consistency}

In addition to spatial and redshift consistency, we impose temporal constraints motivated by plausible physical scenarios linking GRBs and FRBs. These criteria are designed to reflect different evolutionary pathways of the central engine. 
As discussed in Section~\ref{sec:intro}, various physical scenarios predict different time delays between a GRB and a potentially associated FRB. To systematically search for possible connections, we adopt a flexible temporal classification scheme based on the time difference $T_{\rm FRB}-T_{\rm GRB}$.

Specifically, we separate candidate pairs into three groups:
\begin{itemize}
\item \textbf{GRB-before-FRB}: $T_{\rm FRB}-T_{\rm GRB}>2$~yr. This includes scenarios where a long-lived magnetar remnant from either a LGRB or a SGRB produces delayed FRB emission once the surrounding ejecta becomes transparent.
\item \textbf{Intermediate-timescale}: 
$|T_{\rm FRB}-T_{\rm GRB}|<2$~yr. This covers prompt or pre-merger scenarios, e.g., FRB emission from magnetospheric activity shortly before or during a compact binary merger \citep{2014A&AFalcke,2017ApJMetzger}, as well as possible contemporaneous interactions between the GRB jet and its environment \citep{Metzger2019}.
\item \textbf{GRB-after-FRB}: $T_{\rm FRB}-T_{\rm GRB}<-2$~yr. This group is retained for completeness but is not the focus of this work, as the theoretical motivation is less established.
\end{itemize}
The adopted 2~yr threshold is motivated by young-magnetar remnant models, in which FRB emission becomes observable only after the surrounding ejecta and/or magnetar nebula has become sufficiently transparent to radio waves \citep{2014ApJZhang,Yamasaki2018,Dongy2025}. The corresponding transparency timescale is highly uncertain and depends on the ejecta properties, such as, masses and velocities. Representative estimates in the literature include delays from $\sim4$ years \citep{Yamasaki2018} to $\sim10$ years \citep{Margalit2018}, although shorter timescales may also be possible in systems with relatively low-mass ejecta and cleaner environments. We therefore adopt 2~yr as a conservative lower boundary for the delayed-emission regime, so as to avoid excluding potentially relevant candidates. This classification is used primarily for statistical analysis and discussion, and should not be interpreted as a physical constraint on individual candidate pairs.

\subsection{Chance-coincidence Simulation}
\label{sec:method_chance}

To evaluate the probability that the observed associations arise by chance, we perform Monte Carlo simulations that replicate the same selection procedure.

The FRB sample is kept fixed, including their localization maps, event times, and redshift estimation. In each simulation, a mock GRB sample is generated with the same size as the observed sample. The sky positions are drawn isotropically, burst times are sampled from the observed temporal distribution of \textit{Swift} LGRBs, and redshifts are drawn from the expected observed redshift distribution of \citet{Lan2021}. We use their broken-power-law luminosity function and rate evolution, with $L_{\rm c}=10^{53.32}~{\rm erg~s^{-1}}$, $z_{\rm c}=2.33$, $n_1=3.85$, and $n_2=-1.07$. The BAT limit is set to $P_{\rm lim}=1~{\rm ph~cm^{-2}~s^{-1}}$ in the 15--150 keV band. We normalize this $dN/dz$ distribution and sample it by inverse-transform sampling.
For each simulated GRB, we apply the same selection criteria as for the real data, requiring spatial, redshift, and temporal consistency.

We perform 5000 realizations of the simulation. In each realization, we record the number of FRB--GRB pairs that satisfy all criteria, as well as their distribution in localization confidence levels. The statistical significance of the observed associations is then evaluated by comparing with the simulated distributions.

\section{Results}\label{sec:result}
\subsection{The Result for the Matching Procedure}\label{sec:result1}

We first perform spatial matching between \textit{Swift} GRBs and CHIME FRBs with available localization maps. A total of 130 GRB--FRB pairs satisfy the spatial criterion of lying within the FRB localization region with $\mathrm{CL} < 0.68$.

These pairs are classified according to their temporal ordering: 80 pairs with $T_{\rm FRB} - T_{\rm GRB} > 2\,\mathrm{yrs}$, 14 pairs with $T_{\rm GRB} - T_{\rm FRB} > 2\,\mathrm{yrs}$, and 36 pairs with $|T_{\rm GRB} - T_{\rm FRB}| < 2~\mathrm{yr}$. Repeating FRBs are treated as independent bursts in this stage, resulting in a relatively large number of initial matches.

Applying the updated redshift-consistency criterion further reduces the sample to 37 pairs, including 26, 7, and 4 pairs in the three temporal categories, respectively. This reduction highlights both the constraining power of the redshift criterion and the need for more precise redshift measurements in future association studies. 
Among these candidates, 26 belong to the delayed GRB-before-FRB category ($T_{\rm FRB}-T_{\rm GRB}>2$ yr), including 24 FRB--LGRB pairs and two FRB--SGRB pairs associated with sGRB~060502B and sGRB~051221A.
These 26 GRB-before-FRB candidates involve 26 FRBs and 25 distinct GRBs.
Their properties are summarized in Table~\ref{table}.
Short GRBs are labeled with the prefix "sGRB", and repeating FRBs are indicated by "$^r$", with details provided in the table notes. We also checked whether repeaters affect the final sample. Only one candidate pair contains a burst from a known repeater: FRB~20210215C--GRB~130604A, where FRB~20210215C belongs to the repeating source FRB~20181017A. The other bursts from this repeater do not satisfy the spatial criterion for GRB~130604A.
Therefore, treating repeating FRBs as individual bursts rather than as a single source affects at most this one candidate pair.

\begin{figure*}
    \centering
    \includegraphics[width=0.8\linewidth]{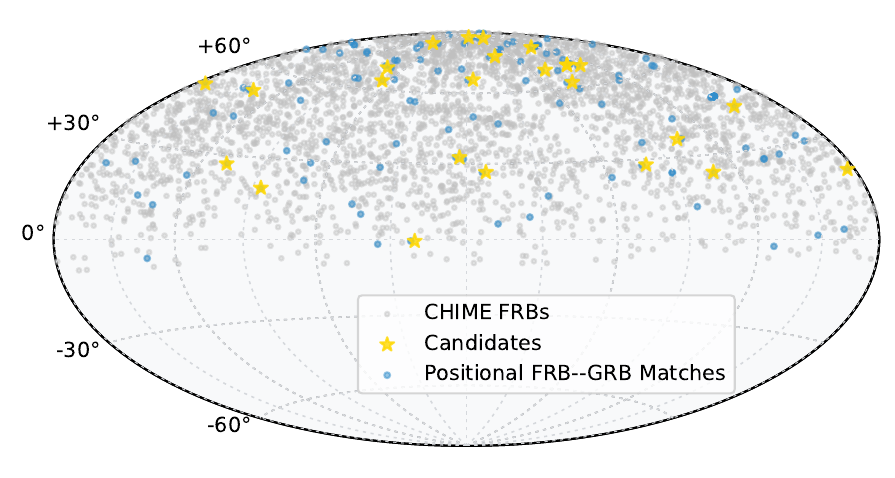}
    \caption{Sky distribution of the FRB--GRB association samples. The gray points represent all FRBs from the second CHIME/FRB catalog. The blue points denote sources that satisfy positional coincidence only, without applying redshift or temporal constraints. The gold star markers indicate 26 redshift-consistent GRB-before-FRB candidates.}
    \label{fig:sky_map}
\end{figure*}

Figure~\ref{fig:sky_map} shows the sky distribution of the candidate pairs. The matches are broadly distributed within the CHIME field of view, with no evidence for significant clustering, indicating an approximately isotropic distribution.

\subsection{Comparison with Monte Carlo Simulations}

\begin{figure}
    \centering
    \includegraphics[width=0.95\linewidth]{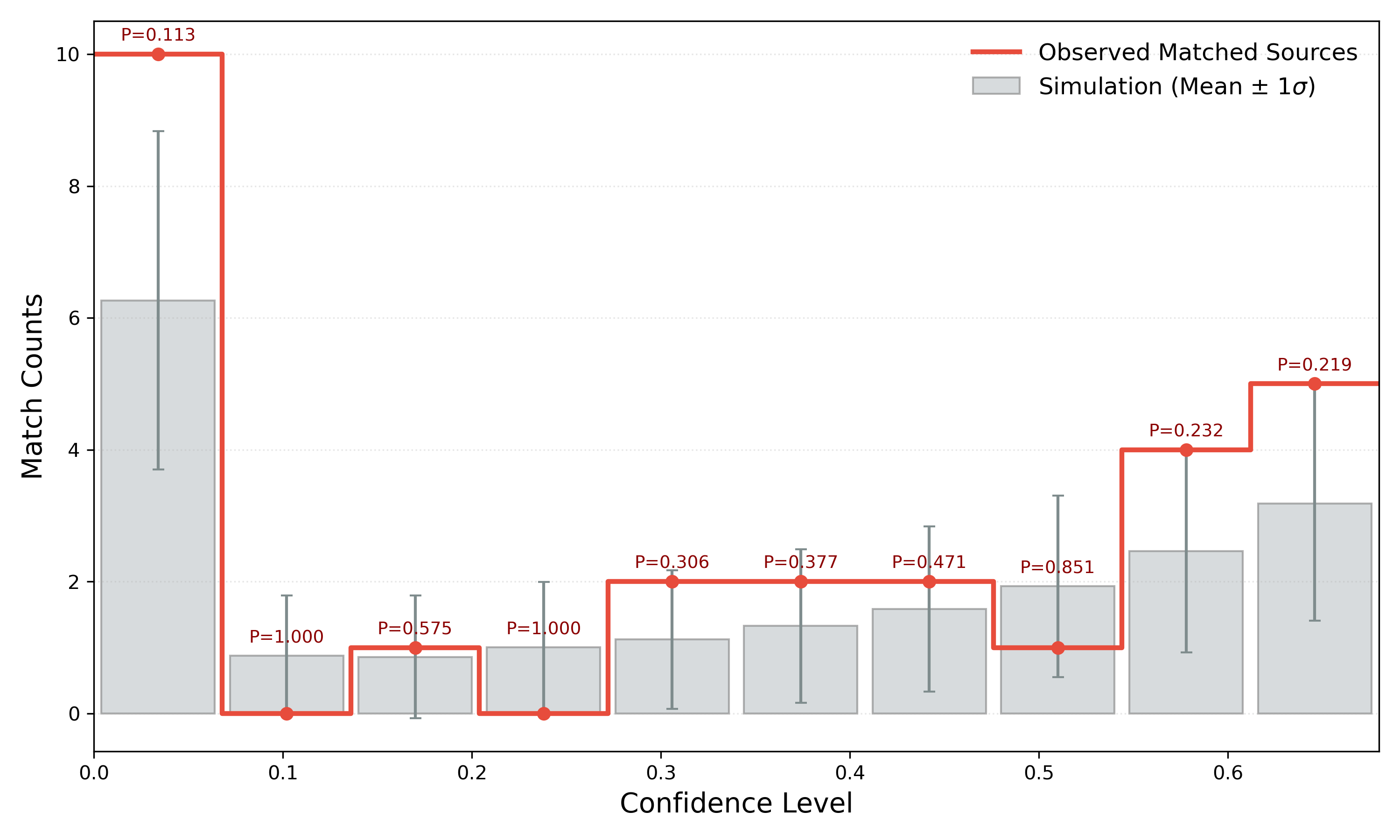}
    \caption{Binned distribution of the confidence level (CL) for matched FRB--GRB pairs. The gray bars represent the mean number of matches in each CL bin derived from simulated GRB samples, with error bars indicating the $1\sigma$ dispersion across realizations. The red step curve and points show the corresponding counts from the real GRB sample. For each bin, the annotated $P$-value denotes the probability that the simulated counts equal or exceed the observed counts, quantifying the statistical significance of the real associations relative to random expectations.}
    \label{fig:results_CL}
\end{figure}

Figure~\ref{fig:results_CL} compares the distribution of CL values evaluated at the GRB positions for the matched GRB--before--FRB pairs with the expectation from 5000 Monte Carlo realizations. The histogram shows the number of matches in each CL bin. Because the CHIME/FRB localization maps are often complex and non-Gaussian, the sky area associated with each CL interval does not follow a simple linear relation with CL. In particular, low-CL intervals can correspond to relatively large sky areas, depending on the detailed structure of the localization map.
The simulated distribution is obtained from the Monte Carlo realizations described in Section~\ref{sec:method_chance}.
In the Monte Carlo simulations, random GRB positions sample the localization maps uniformly in sky area, such that the resulting CL distribution reflects the effective sky area associated with each CL interval. The observed distribution broadly follows the same trend, indicating that the spatial distribution of the matches is consistent with geometric expectations from random coincidences.
Some differences appear in individual bins, but none of them is statistically significant. The smallest probability is $P=0.11$ in the lowest-CL bin, which is still consistent with random fluctuations when several bins are considered.

\section{Discussion}\label{sec:discuss}

\subsection{The Candidates}\label{sec:dis0}

Although the statistical results are consistent with random coincidence, it is still worthwhile to inspect the pairs that satisfy all selection criteria. These pairs satisfy both the spatial criterion from the localization maps and the redshift criterion. The following discussion only shows what these pairs might imply if they are physically related. It should not be taken as evidence for a real association.
 
Among them, two systems are associated with SGRBs, including FRB~20190309A--sGRB~060502B previously reported by \citep{lumx2023}, and FRB~20200227A--sGRB~051221A identified in this work. sGRB~051221A is a short burst with $T_{90}=1.4$~s and $E_{\gamma,\rm iso}\sim10^{51}$~erg at $z=0.546$ \citep{051221Agcn,051221Aredshift}, while the associated FRB has a moderate fluence of $f_{\rm FRB}\sim1.7$~Jy~ms. Under the assumption that the association is physical, we estimate the fluence ratio $\eta = f_{\rm GRB}/f_{\rm FRB}\sim 2\times10^{11}$, where $f_{\rm GRB}$ is the observed gamma-ray fluence and $f_{\rm FRB}$ is the radio fluence converted into energy units using the CHIME bandwidth ($\sim 400~$MHz).
If physically associated, such systems would be broadly consistent with merger-driven scenarios in which FRB activity may precede the neutron star--neutron star binary merger and be suppressed after the merger. Given the temporal configuration adopted here, these cases are treated primarily as illustrative comparisons rather than evidence for such a scenario.

The majority of the sample consists of LGRB-associated candidates, which are more directly relevant to delayed FRB emission scenarios. For instance, FRB~20210917A--GRB~140318A involves a relatively low-energy GRB ($E_{\gamma,\rm iso}\sim8\times10^{50}$~erg) \citep{140318Agcn,140318Aredshift}, yet is followed by a detectable FRB with $f_{\rm FRB}\sim2.0$~Jy~ms after $\sim7.5$~yr. If the association were real, this system would illustrate that delayed FRB activity does not necessarily require an exceptionally energetic prompt event, and may instead depend on the long-term evolution of the central engine.

A contrasting case is FRB~20201111B--GRB~140607A, which is associated with a long-duration GRB ($T_{90}=109.9$~s) with typical fluence $(2.2\pm0.3)\times10^{-6}~\rm erg~cm^{-2}$ \citep{140607Agcn}. The corresponding FRB is non-repeating, with $f_{\rm FRB}\sim1.15$~Jy~ms and $\mathrm{DM}_{\rm exc}\sim261~\rm pc~cm^{-3}$, yielding $\eta\sim5\times10^{11}$ if the association is true. In this case, the GRB exhibits a relatively long duration and typical energy, while the associated FRB shows properties comparable to other events, suggesting that no clear correlation between prompt gamma-ray output and later radio emission would be required.

Across all candidates, the inferred fluence ratios lie in the range $\eta\sim10^{10}$--$10^{11}$. These values are high, but they are not in conflict with the non-detections reported in previous gamma-ray and hard-X-ray searches for FRB counterparts \citep{Yamasaki2016,Cunningham2020,Ridnaia2024}. The large values of $\eta$ imply that, if any of these pairs were physically related, the radio burst would contain only a very small fraction of the high-energy fluence. Since the full sample is consistent with random coincidence, this interpretation remains speculative.

Finally, our results are consistent with those of \citet{Curtin2024}, in the sense that both studies find no significant FRB--GRB association. Our search differs mainly in method: we use the full CHIME localization maps and include redshift information in the candidate selection. Even with these additional constraints, the matches remain consistent with random coincidences.

\subsection{Effects of CHIME Sidelobe Structure on Spatial Matching}\label{sec:dis1}
A key systematic in the spatial association analysis arises from how FRB localization uncertainties are represented. 
To assess this effect, we compared two approaches: (1) a Gaussian approximation based on the central coordinates and 68\% uncertainties reported in the second CHIME/FRB~Catalog, and (2) a map-based method that utilizes the full localization probability distribution.

Under the Gaussian approximation, spatial consistency is evaluated using a normalized elliptical distance assuming a two-dimensional Gaussian error model. In contrast, the map-based approach directly identifies GRB positions that fall within the reported 68\% confidence region of the FRB localization map. Applying the Gaussian prescription yields 30 spatially consistent FRB--GRB pairs. In comparison, the map-based method identifies 130 pairs when considering spatial coincidence alone, indicating a substantial increase in candidate associations.

The catalog-reported positions correspond to the peak response of the synthesized beam and its local uncertainty, but do not capture the full posterior structure of the localization. Owing to CHIME’s cylindrical, East--West interferometric configuration, the beam response can produce multiple sidelobes, leading to spatially extended and multi-modal probability distributions \citep{2019b_chime}. As a result, significant localization probability may reside in regions that are spatially offset from the catalog-reported position.

\begin{figure*}
    \centering
    \includegraphics[width=0.7\linewidth]{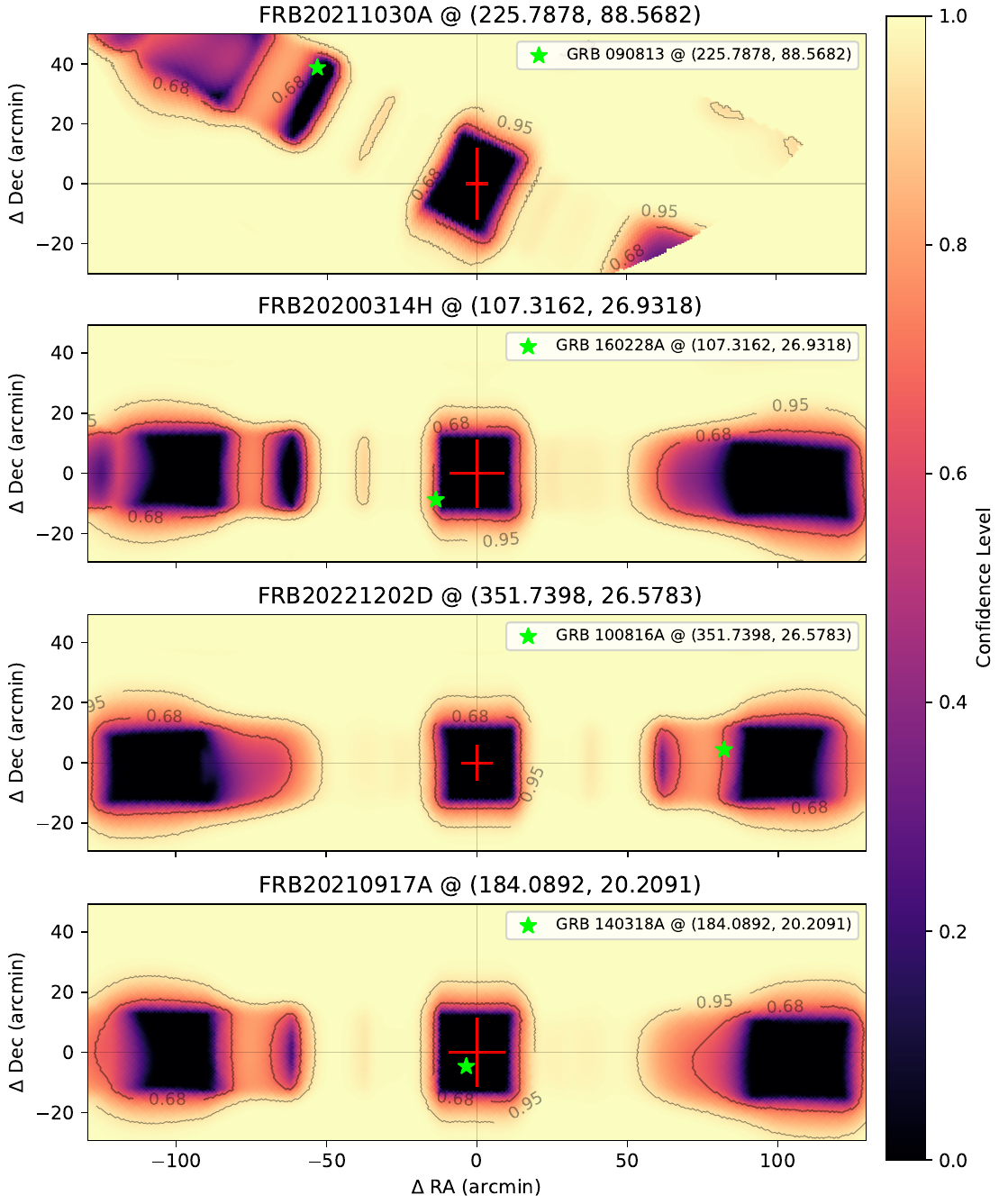}
    \caption{Localization maps of representative FRB–GRB candidate pairs. The red crosses mark the FRB positions reported in the CHIME catalog, while the shaded contours indicate the probabilistic localization regions derived from the FRB beam response, with the 68\% and 95\% confidence levels shown, and the green star symbols denote the positions of the associated GRBs. In the cases of FRB~20211030A--GRB 090813 and FRB~20200314H--GRB 160228A, the GRBs coincide closely with the catalog-reported FRB positions. In contrast, for FRB~20221202D--GRB 100816A and FRB~20210917A--GRB 140318A, the GRBs are offset from the catalog positions and fall into sidelobe regions; however, they remain within the 68\% confidence localization contours of the FRBs. }
    \label{fig:map_example}
\end{figure*}

This effect is illustrated in Figure~\ref{fig:map_example}. In some cases (e.g., FRB~20211030A--GRB~090813 and FRB~20200314H--GRB~160228A), the GRB positions coincide with the catalog-reported FRB locations. However, in other cases (e.g., FRB~20221202D--GRB~100816A and FRB~20210917A--GRB~140318A), the GRBs are clearly offset from the catalog positions and lie within sidelobe-supported probability regions. These regions remain within the 68\% confidence contours but would be excluded under a Gaussian approximation centered on the catalog coordinates.

Therefore, the Gaussian elliptical method systematically underestimates the true localization support and may miss physically plausible associations. The map-based approach, by construction, preserves the full probability distribution and thus provides a more complete and conservative treatment of spatial uncertainties. Although the use of localization maps increases the number of candidate matches, this does not necessarily inflate the statistical significance, as subsequent redshift and temporal constraints substantially reduce the sample. Instead, the map-based method minimizes the risk of excluding genuine associations due to oversimplified localization models. For this reason, it is adopted as the fiducial approach throughout this work.

\subsection{Distribution of Matches in Localization Confidence}\label{sec:dis3}

An interesting feature of our results is that the excess of FRB--GRB associations does not concentrate in the lowest CL bins-corresponding to the highest-probability localization regions—but instead appears at intermediate CL values. 
At first glance, this behavior is consistent with expectations from random spatial coincidence, where no strong preference for the highest-probability regions is present. This is also reflected in the non-significant overall statistical significance obtained from Monte Carlo simulations, suggesting that the majority of identified pairs are likely chance alignments.

The localization uncertainties of GRBs—particularly those without precise afterglow-based positions—can further broaden the effective matching region. When combined with the extended and structured FRB probability maps, this can shift genuine associations toward intermediate CL bins.
If only a small fraction of the sample corresponds to physically associated FRB--GRB pairs, their statistical signature can be easily diluted by a dominant background of random matches. In this regime, even real associations may not produce a strong excess in the highest-probability bins, but instead appear as a mild and distributed enhancement across multiple CL intervals.

Under this interpretation, the observed distribution does not provide strong evidence for a physical association, but it is also not in clear contradiction with scenarios in which a subdominant population of true FRB--GRB pairs is present. Distinguishing between these possibilities will require improved localization precision, larger sample sizes, and more stringent temporal or physical constraints.

\section{Conclusion}\label{sec:con}

We conducted a systematic search for potential associations between CHIME FRBs and \textit{Swift} GRBs using a multi-stage selection procedure incorporating spatial, redshift, and temporal information. By adopting full localization probability maps instead of simplified catalog positions, the analysis accounts for the complex and multi-modal nature of CHIME localizations, resulting in a larger set of spatially consistent pairs. Additional selection criteria based on temporal ordering and redshift consistency were applied to exclude clearly incompatible pairs. These constraints provide a consistent framework for evaluating potential associations while reducing clearly unphysical coincidences.

Monte Carlo simulations indicate that the overall excess of FRB--GRB associations is not statistically significant. The distribution of matches across localization confidence levels is broadly consistent with expectations from random coincidence, although a small contribution from genuine associations cannot be excluded given current uncertainties.
These results suggest that, within the present sample and methodology, there is no compelling statistical evidence for a direct FRB--GRB connection. Future progress will rely on improved localization precision, more uniform sky coverage, and larger samples, which will enable more stringent tests of potential physical links between these two classes of transients.
 
\section*{Acknowledgements}

We thank the anonymous referee for helpful suggestions and comments.
Ye Li thanks Bing Zhang for the helpful discussion.
This work is supported by the National Natural Science Foundation of China (No. 12393813), National Key R\&D Program of China (2024YFA1611704), the Strategic Priority Research Program of the Chinese Academy of Sciences (grant No. XDB0550400),  the China Manned Space Program with grant No. CMS-CSST-2025-A17 and CAS Project for Young Scientists in Basic Research grant YSBR-063.

\software{Astropy~\citep{astropy:2013, astropy:2018, astropy:2022}, {h5py~\citep{h5py_7560547,collette_python_hdf5_2014}}, {skymap~[\url{https://github.com/kadrlica/skymap}]},
{healpy~\citep{healpy_17786647,Zonca2019}}
}

\begin{longtable}{lccccccccc} \label{table} \\
\caption{FRB--GRB matched pairs with redshift estimates.}\\
\hline\hline
FRB & RA & $\sigma_{\rm RA}$ [arcmin] & Dec & $\sigma_{\rm Dec}$ [arcmin] & DM [pc cm$^{-3}$] & $z$ & CL & Match? \\
GRB & RA &  & Dec & $\rm \sigma_{pos}$ [arcmin] & $T_{90}$ &  $z$ &  & \\
\hline\hline
\endfirsthead
\hline\hline
FRB & RA & $\sigma_{\rm RA}$ [arcmin] & Dec & $\sigma_{\rm Dec}$ [arcmin] & DM [pc cm$^{-3}$] & $z$ & CL & Match? \\
GRB & RA &  & Dec & $\rm \sigma_{pos}$  [arcmin] &$T_{90} $  &  $z$ &  & \\
\hline\hline
\endhead
\multicolumn{10}{c}{$ T_{\rm FRB}-T_{\rm GRB}$ ($>$2 yr) (26/80)}\\ \hline
FRB~20180920B & 191.0805 & 13.68 & 63.5156 & 14.28 & 463.1 & $0.34^{+0.11}_{-0.16}$ & 0.000 & N \\
GRB 140909A & 193.6116 &  & 63.5168 & 0.12 & --- & --- &  & \\\hline
FRB~20190309A & 278.9472 & 13.58 & 52.4074 & 14.58 & 356.9 & $0.23^{+0.09}_{-0.11}$ & 0.612 & Y \\
sGRB 060502B & 278.9385 &  & 52.6315 & 0.09 & 0.131 & 0.287 &  & \\\hline
FRB~20190409C & 252.4312 & 12.96 & 71.6140 & 14.37 & 674.6 & $0.54^{+0.12}_{-0.20}$ & 0.473 & N \\
GRB 100316A & 251.9787 &  & 71.8271 & 0.04 & 7.0 & 3.155 &  & \\\hline
FRB~20190429A & 281.0378 & 14.52 & 59.4237 & 14.49 & 470.9 & $0.33^{+0.11}_{-0.16}$ & 0.392 & N \\
GRB 140713A & 281.1059 &  & 59.6335 & 0.02 & 5.30 & $1.92^{+4.07}_{-1.29}$ &  & \\\hline
FRB~20191113B & 169.2738 & 13.39 & 69.1973 & 14.29 & 792.3 & $0.67^{+0.12}_{-0.22}$ & 0.656 & N\\
GRB 080916B & 163.6658 &  & 69.0661 & 0.01 & 32 & $2.87^{+3.32}_{-1.61}$ &  & \\\hline
FRB~20191217B & 239.3967 & 14.61 & 61.8755 & 15.04 & 1428.8 & $1.21^{+0.17}_{-0.26}$ & 0.136 & N \\
GRB 060428B & 235.3570 &  & 62.0248 & 0.02 & 57.900 & 0.348 &  & \\\hline
FRB~20191224B & 185.3263 & 13.76 & 72.1047 & 15.73 & 843.2 & $0.71^{+0.13}_{-0.22}$ & 0.544 & N \\
GRB 150413A & 190.3960 &  & 71.8390 & 1.90 & 263.6 & 3.139 &  & \\\hline
FRB~20200112C & 193.8041 & 13.28 & 30.0619 & 14.71 & 776.4 & $0.66^{+0.13}_{-0.22}$ & 0.359 & N \\
GRB 121211A & 195.5333 &  & 30.1485 & 0.01 & 182 & 1.023 &  & \\\hline
FRB~20200115A & 185.9524 & 16.13 & 87.6369 & 21.09 &  & $0.01^{+0.00}_{-0.00}$ & 0.561 & N\\
GRB 130528A & 139.5051 &  & 87.3012 & 0.03 & 59.4 & $2.54^{+3.95}_{-1.54}$ &  & \\\hline
FRB~20200125C & 176.4086 & 12.78 & 50.2054 & 12.87 & 445.1 & $0.34^{+0.10}_{-0.15}$ & 0.594 & Y \\
GRB 060323 & 174.4379 &  & 49.9853 & 0.03 & 25.400 & $0.32^{+3.00}_{-0.29}$ &  & \\\hline
FRB~20200204K & 262.9962 & 13.21 & 82.1044 & 15.88 & 411.3 & $0.28^{+0.10}_{-0.14}$ & 0.543 & N\\
GRB 131002A & 253.2206 &  & 82.0541 & 0.03 & 55.59 & $3.85^{+3.25}_{-1.97}$ &  & \\\hline
FRB~20200217A & 42.2365 & 10.64 & 9.7406 & 13.49 & 266.3 & $0.15^{+0.07}_{-0.08}$ & 0.000 & N\\
GRB 150819A & 42.3331 &  & 9.8075 & 0.03 & 52.1 & $3.80^{+2.48}_{-1.56}$ &  & \\\hline
FRB~20200222E & 35.0040 & 13.03 & -1.8455 & 21.64 & 560.9 & $0.44^{+0.12}_{-0.18}$ & 0.374 & N \\
GRB 070721B & 33.1373 &  & -2.1946 & 0.02 & 340 & 3.626 &  & \\\hline
FRB~20200224E & 37.5237 & 13.63 & 28.2180 & 15.37 & 303.6 & $0.17^{+0.08}_{-0.09}$ & 0.483 & N\\
GRB 071112C & 39.2122 &  & 28.3713 & 0.01 & 15 & 0.823 &  & \\\hline
FRB~20200227A & 326.6683 & 12.41 & 16.9128 & 14.99 & 613.8 & $0.49^{+0.11}_{-0.19}$ & 0.578 & Y \\
sGRB 051221A & 328.7026 &  & 16.8907 & 0.02 & 1.400 & 0.547 &  & \\\hline
FRB~20200314H & 107.0607 & 14.58 & 27.0790 & 15.80 & 1127.9 & $0.91^{+0.14}_{-0.23}$ & 0.542 & Y\\
GRB 160228A & 107.3162 &  & 26.9318 & 0.03 & 98.36 & $2.16^{+4.03}_{-1.33}$ &  & \\\hline
FRB~20200615D & 243.1663 & 12.86 & 82.1509 & 14.83 & 334.6 & $0.22^{+0.08}_{-0.11}$ & 0.509 & N\\
GRB 131002A & 253.2206 &  & 82.0541 & 0.03 & 55.59 & $3.85^{+3.25}_{-1.97}$ &  & \\\hline
FRB~20200620C & 201.4624 & 13.74 & 61.5316 & 14.81 & 367.4 & $0.25^{+0.10}_{-0.13}$ & 0.589 & N\\
GRB 110402A & 197.4023 &  & 61.2526 & 0.03 & 60.9 & $3.18^{+2.79}_{-1.55}$ &  & \\\hline
FRB~20200722A & 167.4904 & 14.13 & 22.9071 & 15.76 & 220.7 & $0.12^{+0.06}_{-0.06}$ & 0.251 & N \\
GRB 160303A & 168.7007 &  & 22.7420 & 0.03 & 5.0 & $1.60^{+4.14}_{-0.97}$ &  & \\\hline
FRB~20200723A & 265.1491 & 14.00 & 11.8128 & 15.91 & 308.7 & $0.16^{+0.07}_{-0.08}$ & 0.092 & N\\
GRB 170101A & 267.0890 &  & 11.6420 & 1.10 & 2.43 & $3.75^{+3.29}_{-1.97}$ &  & \\\hline
FRB~20200809C & 295.3328 & 13.48 & 86.9662 & 18.13 & 649.3 & $0.50^{+0.13}_{-0.20}$ & 0.333 & Y \\
GRB 150309A & 277.1020 &  & 86.4289 & 0.03 & 242 & $0.78^{+2.69}_{-0.39}$ &  & \\\hline
FRB~20200912A & 21.9233 & 14.78 & -0.6415 & 22.74 & 687.7 & $0.57^{+0.12}_{-0.23}$ & 0.000 & Y\\
GRB 160117A & 20.3674 &  & -0.6554 & 0.03 & 118.58 & $0.32^{+3.00}_{-0.29}$ &  & \\\hline
FRB~20200914A & 230.4868 & 13.41 & 78.4192 & 22.79 & 148.8 & $0.05^{+0.03}_{-0.02}$ & 0.381 & N\\
GRB 060510B & 239.1219 &  & 78.5699 & 0.03 & 275.200 & 4.900 &  & \\\hline
FRB~20200923A & 126.6823 & 13.51 & 73.3324 & 15.46 & 512.1 & $0.39^{+0.10}_{-0.16}$ & 0.546 & N\\
GRB 060515 & 127.2889 &  & 73.5678 & 0.06 & 52.000 & --- &  & \\\hline
FRB~20201003A & 230.0832 & 1.28 & 78.5419 & 11.74 & 516.3 & $0.39^{+0.11}_{-0.17}$ & 0.024 & N\\
GRB 060510B & 239.1219 &  & 78.5699 & 0.03 & 275.200 & 4.900 &  & \\\hline
FRB~20201017D & 27.2954 & 14.16 & 70.2264 & 15.14 & 271.1 & $0.06^{+0.03}_{-0.03}$ & 0.000 & N \\
GRB 180704A & 32.6604 &  & 69.9640 & 0.02 & 19.7 & --- &  & \\\hline
FRB~20201017B & 65.9996 & 15.41 & 64.1672 & 15.97 & 333.3 & $0.13^{+0.07}_{-0.07}$ & 0.000 & Y \\
GRB 120106A & 66.1075 &  & 64.0384 & 0.02 & 61.6 & $0.32^{+3.00}_{-0.29}$ &  & \\\hline
FRB~20201111B & 86.3465 & 13.43 & 18.8230 & 15.39 & 424.2 & $0.19^{+0.08}_{-0.10}$ & 0.000 & Y\\
GRB 140607A & 86.3730 &  & 18.9040 & 1.90 & 109.9 & $0.32^{+3.00}_{-0.29}$ &  & \\\hline
FRB~20201127B & 252.2604 & 13.51 & 75.0855 & 23.94 & 672.4 & $0.55^{+0.11}_{-0.21}$ & 0.000 & N\\
GRB 180812A & 245.8352 &  & 74.6647 & 0.01 & 16.51 & $4.07^{+3.04}_{-1.90}$ &  & \\\hline
FRB~20201129B & 90.3778 & 11.51 & 79.5606 & 16.81 & 303.3 & $0.17^{+0.08}_{-0.09}$ & 0.400 & N \\
GRB 050410 & 89.8088 &  & 79.6034 & 0.03 & 42.500 & --- &  & \\\hline
FRB~20201202C & 80.0328 & 14.84 & 73.2307 & 16.16 & 962.9 & $0.78^{+0.14}_{-0.23}$ & 0.000 & N\\
GRB 041226 & 79.6890 &  & 73.3290 & 3.00 & 89.700 & 2.230 &  & \\\hline
FRB~20201206D & 276.2605 & 13.39 & 36.4041 & 14.37 & 354.3 & $0.21^{+0.09}_{-0.11}$ & 0.356 & N\\
GRB 080325 & 277.8930 &  & 36.5239 & 0.03 & 128.4 & 2.000 &  & \\\hline
FRB~20201207F & 238.2665 & 14.50 & 74.6057 & 16.08 & 1345.0 & $1.13^{+0.17}_{-0.27}$ & 0.211 & N\\
GRB 180812A & 245.8352 &  & 74.6647 & 0.01 & 16.51 & $4.07^{+3.04}_{-1.90}$ &  & \\\hline
FRB~20201231A & 201.4703 & 12.59 & 61.1427 & 12.92 & 682.7 & $0.57^{+0.12}_{-0.21}$ & 0.475 & N\\
GRB 110402A & 197.4023 &  & 61.2526 & 0.03 & 60.9 & $3.18^{+2.79}_{-1.55}$ &  & \\\hline
FRB~20210101E & 356.3604 & 9.74 & 49.5958 & 9.65 & 860.8 & $0.65^{+0.12}_{-0.21}$ & 0.549 & N\\
GRB 100212A & 356.4183 &  & 49.4943 & 0.03 & 136 & $3.25^{+3.23}_{-1.73}$ &  & \\\hline
FRB~20210105C & 211.8952 & 10.96 & 1.3180 & 15.99 & 291.1 & $0.19^{+0.08}_{-0.10}$ & 0.000 & N\\
GRB 140102A & 211.9194 &  & 1.3333 & 0.01 & 65 & $2.64^{+3.91}_{-1.61}$ &  & \\\hline
FRB~20210131B & 328.5821 & 13.46 & 77.0434 & 15.78 & 438.7 & $0.28^{+0.10}_{-0.13}$ & 0.000 & Y\\
GRB 050713A & 320.5387 &  & 77.0747 & 0.02 & 124.700 & $0.32^{+3.00}_{-0.29}$ &  & \\\hline
FRB~20210215C$^r$ & 252.9044 & 13.70 & 68.2984 & 15.42 &  & $1.07^{+0.17}_{-0.26}$ & 0.458 & Y \\
GRB 130604A & 250.1870 &  & 68.2264 & 0.03 & 37.7 & 1.060 &  & \\\hline
FRB~20210216G & 83.3544 & 15.06 & 73.2368 & 16.58 & 957.3 & $0.79^{+0.13}_{-0.22}$ & 0.437 & N \\
GRB 041226 & 79.6890 &  & 73.3290 & 3.00 & 89.700 & 2.230 &  & \\\hline
FRB~20210315A & 246.8502 & 11.83 & 78.5119 & 13.62 & 722.2 & $0.58^{+0.13}_{-0.21}$ & 0.517 & N\\
GRB 060510B & 239.1219 &  & 78.5699 & 0.03 & 275.200 & 4.900 &  & \\\hline
FRB~20210327D & 73.8467 & 13.38 & 83.1202 & 20.15 & 864.2 & $0.69^{+0.13}_{-0.24}$ & 0.020 & Y\\
GRB 130603A & 86.8912 &  & 82.9087 & 0.01 & 76 & $0.32^{+3.00}_{-0.29}$ &  & \\\hline
FRB~20210412C & 68.9466 & 12.99 & 22.3130 & 15.14 & 463.8 & $0.29^{+0.10}_{-0.14}$ & 0.627 & N\\
GRB 090712 & 70.0970 &  & 22.5250 & 1.60 & 145 & $2.88^{+2.01}_{-1.18}$ &  & \\\hline
FRB~20210424B & 4.9459 & 14.57 & 32.6932 & 14.89 & 774.5 & $0.63^{+0.13}_{-0.21}$ & 0.000 & Y \\
GRB 080515 & 3.1660 &  & 32.5640 & 1.60 & 21 & $0.49^{+1.99}_{-0.31}$ &  & \\\hline
FRB~20210504C & 238.0948 & 12.61 & 78.5053 & 14.28 & 719.5 & $0.59^{+0.12}_{-0.22}$ & 0.645 & Y\\
GRB 181213A & 248.2703 &  & 78.4969 & 0.01 & 15.31 & $0.32^{+3.00}_{-0.29}$ &  & \\\hline
FRB~20210603B & 122.8461 & 13.52 & 21.9857 & 16.02 & 747.1 & $0.59^{+0.13}_{-0.22}$ & 0.237 & N\\
GRB 141121A & 122.6693 &  & 22.2173 & 0.01 & 549.9 & 1.470 &  & \\\hline
FRB~20210609E & 245.7006 & 13.41 & 60.2837 & 14.28 & 894.0 & $0.75^{+0.14}_{-0.21}$ & 0.430 & N \\
GRB 081025 & 245.2980 &  & 60.4660 & 2.87 & 23 & $3.26^{+2.96}_{-1.62}$ &  & \\\hline
FRB~20210612C & 9.0792 & 13.76 & 44.1608 & 14.58 & 181.8 & $0.05^{+0.03}_{-0.02}$ & 0.512 & N \\
GRB 150530B & 7.4960 &  & 44.2900 & 3.00 &  & $3.40^{+3.53}_{-1.93}$ &  & \\\hline
FRB~20210805B & 255.7080 & 12.04 & 36.4470 & 12.41 & 1514.0 & $1.27^{+0.19}_{-0.27}$ & 0.375 & Y\\
GRB 110928A & 257.7327 &  & 36.5357 & 0.03 & 26.7 & $2.10^{+4.13}_{-1.35}$ &  & \\\hline
FRB~20210827A & 140.1157 & 9.29 & 43.1988 & 9.59 & 266.2 & $0.16^{+0.07}_{-0.08}$ & 0.455 & N \\
GRB 161017A & 142.7692 &  & 43.1268 & 0.01 & 216.3 & 2.010 &  & \\\hline
FRB~20210901B & 69.0364 & 15.02 & 29.0841 & 16.25 & 2044.0 & $1.65^{+0.20}_{-0.31}$ & 0.174 & N \\
GRB 131227A & 67.3782 &  & 28.8830 & 0.02 & 18.0 & 5.300 &  & \\\hline
FRB~20210917A & 184.0263 & 14.52 & 20.2869 & 16.37 & 1078.0 & $0.94^{+0.14}_{-0.23}$ & 0.000 & Y \\
GRB 140318A & 184.0892 &  & 20.2091 & 0.03 & 8.43 & 1.020 &  & \\\hline
FRB~20210927A & 290.9223 & 3.99 & 68.6305 & 15.84 & 329.5 & $0.20^{+0.08}_{-0.10}$ & 0.615 & Y \\
GRB 080503 & 286.6194 &  & 68.7931 & 0.03 & 170 & $0.32^{+3.00}_{-0.29}$ &  & \\\hline
FRB~20211008F & 99.2417 & 14.29 & 81.1764 & 17.70 & 1033.4 & $0.86^{+0.14}_{-0.24}$ & 0.140 & N \\
GRB 110801A & 89.4370 &  & 80.9559 & 0.01 & 385 & 1.858 &  & \\\hline
FRB~20211011C & 101.7919 & 14.23 & 52.4483 & 14.11 & 908.0 & $0.73^{+0.14}_{-0.23}$ & 0.000 & N\\
GRB 190515B & 98.8484 &  & 52.3173 & 0.03 & 46.4 & $3.18^{+1.36}_{-0.97}$ &  & \\\hline
FRB~20211030A$^{r}$ & 187.3746 & 17.95 & 88.2313 & 19.96 & 330.3 & $0.20^{+0.08}_{-0.10}$ & 0.082 & N\\
GRB 090813 & 225.7878 &  & 88.5682 & 0.02 & 7.1 & $2.58^{+3.87}_{-1.60}$ &  & \\\hline
FRB~20211105C & 138.0583 & 12.50 & 87.3581 & 14.22 & 321.9 & $0.19^{+0.08}_{-0.10}$ & 0.000 & N\\
GRB 130528A & 139.5051 &  & 87.3012 & 0.03 & 59.4 & $2.54^{+3.95}_{-1.54}$ &  & \\\hline
FRB~20211127E & 256.3828 & 14.12 & 69.5364 & 16.14 & 840.1 & $0.70^{+0.13}_{-0.22}$ & 0.039 & Y \\
GRB 070219 & 260.1919 &  & 69.3702 & 0.03 & 16.600 & $0.32^{+3.00}_{-0.29}$ &  & \\\hline
FRB~20211201B & 206.4291 & 13.73 & 43.9970 & 14.74 & 447.1 & $0.35^{+0.11}_{-0.16}$ & 0.000 & Y\\
GRB 080319A & 206.3332 &  & 44.0800 & 0.03 & 64 & $0.32^{+3.00}_{-0.29}$ &  & \\\hline
FRB~20220104B & 31.9555 & 13.00 & 55.8235 & 13.44 & 294.6 & $0.06^{+0.03}_{-0.03}$ & 0.115 & N\\
GRB 100906A & 28.6838 &  & 55.6304 & 0.01 & 114.4 & 1.727 &  & \\\hline
FRB~20220303C & 142.2714 & 13.86 & 52.5217 & 14.95 & 778.9 & $0.65^{+0.13}_{-0.21}$ & 0.328 & Y \\
GRB 171004A & 139.1694 &  & 52.6931 & 0.02 & 107.1 & $1.03^{+3.42}_{-0.56}$ &  & \\\hline
FRB~20220426D & 195.0640 & 13.22 & 32.1273 & 14.59 & 188.0 & $0.10^{+0.05}_{-0.05}$ & 0.000 & N\\
GRB 141220A & 195.0657 &  & 32.1464 & 0.01 & 7.21 & 1.319 &  & \\\hline
FRB~20220623D & 255.9685 & 12.61 & 69.5308 & 13.83 & 842.6 & $0.70^{+0.14}_{-0.22}$ & 0.608 & Y\\
GRB 070219 & 260.1919 &  & 69.3702 & 0.03 & 16.600 & $0.32^{+3.00}_{-0.29}$ &  & \\\hline
FRB~20220726B & 73.6794 & 11.79 & 69.8839 & 12.43 & 686.9 & $0.51^{+0.12}_{-0.20}$ & 0.657 & Y\\
GRB 060124 & 77.1077 &  & 69.7407 & 0.02 & ~750 & $0.32^{+3.00}_{-0.29}$ &  & \\\hline
FRB~20220809A & 282.6572 & 11.99 & 27.8601 & 12.92 & 792.4 & $0.58^{+0.12}_{-0.22}$ & 0.011 & Y\\
GRB 181202A & 280.7349 &  & 27.9597 & 0.01 & 6.56 & $0.32^{+3.00}_{-0.29}$ &  & \\\hline
FRB~20220904F$^r$ & 46.2606 & 14.20 & 13.6330 & 18.13 & 187.3 & $0.08^{+0.04}_{-0.04}$ & 0.587 & N\\
GRB 120802A & 44.8426 &  & 13.7683 & 0.02 & 50 & 3.796 &  & \\\hline
FRB~20221015A & 313.2241 & 15.11 & 87.7235 & 19.40 & 554.0 & $0.41^{+0.12}_{-0.17}$ & 0.356 & Y \\
GRB 110223A & 345.8522 &  & 87.5579 & 0.03 & 7.0 & $0.32^{+3.00}_{-0.29}$ &  & \\\hline
FRB~20221109D & 131.1778 & 13.00 & 73.7720 & 14.69 & 2170.7 & $1.78^{+0.23}_{-0.32}$ & 0.536 & N\\
GRB 060515 & 127.2889 &  & 73.5678 & 0.06 & 52.000 & --- &  & \\\hline
FRB~20221130C & 227.6540 & 13.39 & 30.8769 & 13.72 & 723.3 & $0.61^{+0.13}_{-0.21}$ & 0.321 & N \\
GRB 181027A & 225.3610 &  & 30.9370 & 1.10 & 81.16 & $3.94^{+3.04}_{-1.88}$ &  & \\\hline
FRB~20221202D & 353.2679 & 14.58 & 26.5140 & 15.21 & 957.9 & $0.81^{+0.14}_{-0.23}$ & 0.655 & Y \\
GRB 100816A & 351.7398 &  & 26.5783 & 0.01 & 2.9 & 0.803 &  & \\\hline
FRB~20221209B & 189.7913 & 12.26 & 78.5873 & 14.10 & 398.4 & $0.28^{+0.10}_{-0.14}$ & 0.514 & N\\
GRB 160227A & 194.8075 &  & 78.6792 & 0.01 & 316.5 & 2.380 &  & \\\hline
FRB~20221210F & 347.5532 & 13.71 & 6.0833 & 19.94 & 251.3 & $0.14^{+0.07}_{-0.08}$ & 0.086 & N \\
GRB 110119A & 348.5859 &  & 5.9863 & 0.01 & 208 & $3.21^{+3.45}_{-1.79}$ &  & \\\hline
FRB~20230228B & 303.1613 & 14.59 & 54.2790 & 14.69 & 662.1 & $0.46^{+0.12}_{-0.17}$ & 0.000 & N\\
GRB 121128A & 300.6000 &  & 54.2998 & 0.01 & 23.3 & 2.200 &  & \\\hline
FRB~20230307E & 251.5923 & 12.82 & 23.6971 & 14.63 & 1735.8 & $1.44^{+0.20}_{-0.29}$ & 0.152 & Y\\
GRB 120312A & 251.7882 &  & 23.8582 & 0.03 & 14.2 & $2.50^{+3.78}_{-1.60}$ &  & \\\hline
FRB~20230316H & 358.7996 & 2.78 & 66.1087 & 14.17 & 496.1 & $0.21^{+0.08}_{-0.10}$ & 0.654 & Y \\
GRB 070704 & 354.6977 &  & 66.2532 & 0.03 & 380 & $0.32^{+3.00}_{-0.29}$ &  & \\\hline
FRB~20230323C & 213.5683 & 13.90 & 27.5705 & 14.76 & 896.9 & $0.78^{+0.13}_{-0.23}$ & 0.000 & N \\
GRB 060204B & 211.8125 &  & 27.6771 & 0.02 & 139.400 & $2.60^{+3.88}_{-1.66}$ &  & \\\hline
FRB~20230405E & 33.1337 & 15.73 & 37.8995 & 15.91 & 745.0 & $0.60^{+0.12}_{-0.20}$ & 0.202 & N \\
GRB 190109A & 33.2023 &  & 38.1082 & 0.02 & 115.0 & $3.12^{+3.27}_{-1.69}$ &  & \\\hline
FRB~20230410D & 283.9426 & 13.28 & 62.5839 & 13.37 & 1206.4 & $1.00^{+0.15}_{-0.26}$ & 0.469 & N\\
sGRB 160821B & 279.9762 &  & 62.3914 & 0.04 & 0.48 & 0.162 &  & \\\hline
FRB~20230428C & 312.8761 & 14.60 & 64.9336 & 15.18 & 456.4 & $0.28^{+0.09}_{-0.14}$ & 0.191 & N \\
GRB 090727 & 315.9608 &  & 64.9248 & 0.03 & 302 & --- &  & \\\hline
FRB~20230429A & 126.8752 & 10.81 & 43.4291 & 3.10 & 227.6 & $0.12^{+0.05}_{-0.06}$ & 0.649 & N\\
GRB 100213B & 124.2823 &  & 43.4477 & 0.06 & 48.0 & --- &  & \\\hline
FRB~20230501D & 132.6145 & 14.08 & 11.4446 & 18.76 &  & $7.75^{+0.75}_{-1.07}$ & 0.513 & N\\
GRB 050416B & 133.8480 &  & 11.1860 & 1.30 & 3.400 & --- &  & \\\hline

\multicolumn{10}{c}{$ |T_{\rm GRB}-T_{\rm FRB}|<2\,\mathrm{yr}$(4/36)} \\ \hline

FRB~20181017B & 237.4245 & 13.75 & 78.4956 & 15.13 & 306.9 & $0.19^{+0.08}_{-0.09}$ & 0.533 & N\\
GRB 200925B & 246.7858 &  & 78.3907 & 0.02 & 18.25 & $3.49^{+3.28}_{-1.83}$ &  & \\\hline
FRB~20181030B$^r$ & 164.2155 & 13.50 & 73.7395 & 25.46 & 103.5 & $0.01^{+0.01}_{-0.00}$ & 0.006 & N\\
GRB 180618A & 169.9410 &  & 73.8371 & 0.01 & 47.4 & $2.59^{+1.44}_{-0.92}$ &  & \\\hline
FRB~20181030A$^r$ & 163.4206 & 13.24 & 73.7400 & 24.63 & 103.6 & $0.01^{+0.01}_{-0.00}$ & 0.228 & N\\
GRB 180618A & 169.9410 &  & 73.8371 & 0.01 & 47.4 & $2.59^{+1.44}_{-0.92}$ &  & \\\hline
FRB~20181117A & 147.7138 & 12.96 & 52.5121 & 12.82 & 959.1 & $0.81^{+0.14}_{-0.23}$ & 0.000 & N \\
GRB 161217A & 150.6510 &  & 52.3590 & 0.03 & 18.2 & $4.12^{+2.95}_{-1.87}$ &  & \\\hline
FRB~20190303A$^r$ & 208.0760 & 14.20 & 48.2524 & 13.77 & 222.0 & $0.13^{+0.06}_{-0.07}$ & 0.000 & N\\
GRB 201229A & 210.6892 &  & 48.1981 & 0.01 & 53.3 & $3.12^{+3.37}_{-1.72}$ &  & \\\hline
FRB~20190304C & 223.0739 & 14.02 & 26.7173 & 14.81 & 565.0 & $0.46^{+0.12}_{-0.20}$ & 0.000 & N \\
GRB 201128A & 223.1076 &  & 26.6848 & 0.03 & 5.41 & $3.67^{+3.06}_{-1.80}$ &  & \\\hline
FRB~20190421A$^r$ & 208.1128 & 13.91 & 48.2469 & 13.33 & 223.0 & $0.13^{+0.06}_{-0.07}$ & 0.059 & N\\
GRB 201229A & 210.6892 &  & 48.1981 & 0.01 & 53.3 & $3.12^{+3.37}_{-1.72}$ &  & \\\hline
FRB~20190702B$^r$ & 208.1480 & 11.51 & 48.2526 & 11.42 & 222.1 & $0.13^{+0.06}_{-0.07}$ & 0.045 & N \\
GRB 201229A & 210.6892 &  & 48.1981 & 0.01 & 53.3 & $3.12^{+3.37}_{-1.72}$ &  & \\\hline
FRB~20190714D & 298.1368 & 13.56 & 23.6043 & 15.78 & 1983.9 & $1.32^{+0.18}_{-0.28}$ & 0.648 & Y\\
GRB 180314B & 297.8867 &  & 23.6241 & 0.03 & 73.0 & $1.02^{+3.10}_{-0.54}$ &  & \\\hline
FRB~20190808A & 35.4679 & 14.43 & 56.2269 & 14.32 & 793.0 & $0.53^{+0.12}_{-0.18}$ & 0.471 & N \\
GRB 201228B & 35.5940 &  & 56.0150 & 3.00 & nan & $2.81^{+3.60}_{-1.66}$ &  & \\\hline
FRB~20191010A & 175.4534 & 13.45 & 73.7567 & 25.44 & 1218.2 & $1.04^{+0.15}_{-0.25}$ & 0.435 & N \\
GRB 180618A & 169.9410 &  & 73.8371 & 0.01 & 47.4 & $2.59^{+1.44}_{-0.92}$ &  & \\\hline
FRB~20191117A$^r$ & 210.7351 & 12.85 & 48.2509 & 13.33 & 221.3 & $0.13^{+0.06}_{-0.07}$ & 0.000 & N\\
GRB 201229A & 210.6892 &  & 48.1981 & 0.01 & 53.3 & $3.12^{+3.37}_{-1.72}$ &  & \\\hline
FRB~20191215A$^r$ & 208.2523 & 12.26 & 48.2550 & 12.11 & 222.5 & $0.13^{+0.06}_{-0.07}$ & 0.013 & N\\
GRB 201229A & 210.6892 &  & 48.1981 & 0.01 & 53.3 & $3.12^{+3.37}_{-1.72}$ &  & \\\hline
FRB~20200112A$^r$ & 207.7768 & 12.75 & 48.2541 & 13.18 & 221.4 & $0.13^{+0.06}_{-0.07}$ & 0.665 & N\\
GRB 201229A & 210.6892 &  & 48.1981 & 0.01 & 53.3 & $3.12^{+3.37}_{-1.72}$ &  & \\\hline
FRB~20200229C & 179.4406 & 14.50 & 67.1723 & 15.14 & 1753.3 & $1.46^{+0.20}_{-0.30}$ & 0.598 & Y\\
GRB 190613A & 182.5292 &  & 67.2353 & 0.01 & 17.6 & $3.39^{+3.52}_{-1.89}$ &  & \\\hline
FRB~20200613C$^r$ & 210.4026 & 14.93 & 48.2555 & 15.19 & 222.5 & $0.13^{+0.06}_{-0.07}$ & 0.274 & N\\
GRB 201229A & 210.6892 &  & 48.1981 & 0.01 & 53.3 & $3.12^{+3.37}_{-1.72}$ &  & \\\hline
FRB~20200809G$^r$ & 208.2540 & 11.42 & 48.2604 & 11.45 & 222.0 & $0.13^{+0.06}_{-0.07}$ & 0.478 & N\\
GRB 201229A & 210.6892 &  & 48.1981 & 0.01 & 53.3 & $3.12^{+3.37}_{-1.72}$ &  & \\\hline
FRB~20200829A$^r$ & 208.0164 & 14.47 & 48.2594 & 15.53 & 222.1 & $0.13^{+0.06}_{-0.07}$ & 0.000 & N\\
GRB 201229A & 210.6892 &  & 48.1981 & 0.01 & 53.3 & $3.12^{+3.37}_{-1.72}$ &  & \\\hline
FRB~20200924A & 223.3128 & 13.55 & 26.7255 & 15.33 & 1835.3 & $1.55^{+0.20}_{-0.27}$ & 0.165 & N \\
GRB 201128A & 223.1076 &  & 26.6848 & 0.03 & 5.41 & $3.67^{+3.06}_{-1.80}$ &  & \\\hline
FRB~20201016A & 126.7916 & 2.22 & 57.5561 & 13.68 & 421.1 & $0.30^{+0.10}_{-0.14}$ & 0.238 & N\\
GRB 210226A & 124.1153 &  & 57.5839 & 0.03 & 20.80 & $3.97^{+3.03}_{-1.86}$ &  & \\\hline
FRB~20201103A$^r$ & 208.0070 & 13.58 & 48.2628 & 14.35 & 222.3 & $0.13^{+0.06}_{-0.07}$ & 0.000 & N \\
GRB 201229A & 210.6892 &  & 48.1981 & 0.01 & 53.3 & $3.12^{+3.37}_{-1.72}$ &  & \\\hline
FRB~20210302C$^r$ & 207.9115 & 12.57 & 48.2602 & 12.95 & 222.0 & $0.13^{+0.06}_{-0.07}$ & 0.465 & N\\
GRB 201229A & 210.6892 &  & 48.1981 & 0.01 & 53.3 & $3.12^{+3.37}_{-1.72}$ &  & \\\hline
FRB~20210315A & 246.8502 & 11.83 & 78.5119 & 13.62 & 722.2 & $0.58^{+0.13}_{-0.21}$ & 0.000 & N\\
GRB 200925B & 246.7858 &  & 78.3907 & 0.02 & 18.25 & $3.49^{+3.28}_{-1.83}$ &  & \\\hline
FRB~20210425A & 57.7221 & 14.10 & 64.9574 & 15.48 & 932.3 & $0.69^{+0.14}_{-0.20}$ & 0.438 & Y \\
sGRB 201006A & 61.8925 &  & 65.1646 & 0.03 & 0.49 & $0.58^{+0.06}_{-0.06}$ &  & \\\hline
FRB~20210504C & 238.0948 & 12.61 & 78.5053 & 14.28 & 719.5 & $0.59^{+0.12}_{-0.22}$ & 0.607 & N \\
GRB 200925B & 246.7858 &  & 78.3907 & 0.02 & 18.25 & $3.49^{+3.28}_{-1.83}$ &  & \\\hline
FRB~20210514C$^r$ & 207.9595 & 14.92 & 48.2558 & 15.09 & 222.0 & $0.13^{+0.06}_{-0.07}$ & 0.083 & N\\
GRB 201229A & 210.6892 &  & 48.1981 & 0.01 & 53.3 & $3.12^{+3.37}_{-1.72}$ &  & \\\hline
FRB~20210710B & 249.6799 & 13.72 & 43.9296 & 14.73 & 2283.1 & $1.89^{+0.23}_{-0.35}$ & 0.228 & Y\\
GRB 191101A & 251.8372 &  & 43.7407 & 0.02 & 137.95 & $0.32^{+3.00}_{-0.29}$ &  & \\\hline
FRB~20210911A & 152.6946 & 13.32 & 24.6306 & 15.26 & 348.7 & $0.23^{+0.09}_{-0.12}$ & 0.000 & N\\
GRB 211024B & 154.7127 &  & 24.5682 & 0.03 & 603.5 & $3.20^{+2.34}_{-1.35}$ &  & \\\hline
FRB~20211014B$^r$ & 207.9208 & 12.60 & 48.2662 & 12.76 & 223.0 & $0.13^{+0.06}_{-0.07}$ & 0.138 & N\\
GRB 201229A & 210.6892 &  & 48.1981 & 0.01 & 53.3 & $3.12^{+3.37}_{-1.72}$ &  & \\\hline
FRB~20211104F & 6.1359 & 15.14 & 84.1887 & 22.77 & nan & $1.08^{+0.16}_{-0.24}$ & 0.000 & N\\
GRB 231104A & 23.8068 &  & 83.7935 & 0.03 & 46.77 & $3.30^{+3.07}_{-1.69}$ &  & \\\hline
FRB~20211112E & 81.9229 & 14.61 & 70.3123 & 15.70 & 741.1 & $0.59^{+0.12}_{-0.19}$ & 0.074 & N \\
GRB 210321A & 87.8948 &  & 70.1305 & 0.01 & 8.21 & 1.487 &  & \\\hline
FRB~20211125D & 276.8685 & 13.03 & 67.8301 & 13.31 & 532.0 & $0.41^{+0.10}_{-0.16}$ & 0.028 & Y\\
GRB 200906A & 272.2754 &  & 67.8801 & 0.03 & 70.90 & $0.52^{+0.93}_{-0.23}$ &  & \\\hline
FRB~20220103C$^r$ & 208.0501 & 14.25 & 48.2593 & 14.14 & 221.4 & $0.13^{+0.06}_{-0.07}$ & 0.141 & N\\
GRB 201229A & 210.6892 &  & 48.1981 & 0.01 & 53.3 & $3.12^{+3.37}_{-1.72}$ &  & \\\hline
FRB~20220629A & 298.0590 & 13.64 & 80.2045 & 16.86 & 571.6 & $0.44^{+0.11}_{-0.17}$ & 0.000 & N\\
GRB 230328B & 291.0078 &  & 80.0096 & 0.01 & 19.23 & 0.090 &  & \\\hline
FRB~20221116B & 334.8568 & 8.85 & 8.6964 & 14.15 & 651.9 & $0.54^{+0.12}_{-0.20}$ & 0.635 & N\\
GRB 240511A & 336.6758 &  & 8.5128 & 0.03 & 148.55 & $1.96^{+4.21}_{-1.25}$ &  & \\\hline
FRB~20230915C & 1.3256 & 14.00 & 31.8783 & 13.95 & 848.5 & $0.69^{+0.14}_{-0.22}$ & 0.000 & N\\
GRB 220101A & 1.3533 &  & 31.7690 & 0.01 & 173.36 & 4.610 &  & \\\hline

\end{longtable}

\tablecomments{$^r$ FRB~20211030A comes from repeater FRB~20211030A; FRB~20210215C comes from repeater FRB~20181017A; FRB~20220904F comes from repeater FRB~20220618C; FRB~20190303A, FRB~20190421A, FRB~20190702B, FRB~20191215A, FRB~20191117A, FRB~20200112A, FRB~20200613C, FRB~20200809G, FRB~20200829A, FRB~20201103A, FRB~20210302C, FRB~20210514C, FRB~20211014B, FRB~20220103C come from repeater FRB~20190303A; FRB~20181030A and FRB~20181030B come from repeater FRB~20181030A}

\bibliography{sample631}{}
\bibliographystyle{aasjournalv7}

\end{document}